# TIME SCALES FOR COLD-WELDING AND THE ORIGINS OF STICK-SLIP FRICTION

Raffi Budakian[†], Seth J. Putterman

Physics Department University of California, Los Angeles, CA 90095

*Abstract:* Data from an instrument that measures conductance, stiffness and rupture strength of junctions indicates that when two macroscopic metal surfaces are brought into contact a nanometer size junction spontaneously forms over a long time scale (~ 60 s). Furthermore, the parameters of junction rupture match the observed dynamics of stick-slip friction. This suggests that stick-slip friction has its origins in the formation and rupture of junctions that form between surfaces in contact.

[†] To whom correspondence should be addressed; electronic mail: budak@physics.ucla.edu

The microscopic origin of the force of friction between surfaces in relative motion has been interpreted in terms of models that invoke asperities [1-3], bond formation [4,5], impurities [6], collective elastic deformations [7], interactions between electronic states [8], and charge separation [9,10]. The electrical and mechanical properties of nanometer-sized metal junctions have drawn considerable interest in recent years for their potential technological applications as well as for the light they shed on fundamental questions related to bond formation, friction, wear, and fracture between metals. Most experiments that study junction formation do so by bringing the sharp tip of a scanning tunneling microscope (STM) into contact with a surface[11-16]. In order to investigate the role of bonding on friction between macroscopic metal bodies, we have measured the dynamics of cold-welds that spontaneously form between surfaces brought into contact. As shown in Figure 1 we find that upon contact the two surfaces 'suck-in' through the formation of a junction whose initial angstrom size radius grows by a factor of 50 in about a minute. Furthermore the forces of stick-slip friction between these macroscopic bodies match the forces required to break the nanometer size junctions.

The experiment is performed with a fiber-optic cantilever surface probe microscope [Figure 2] that can measure displacement and conductance accompanying the sub-nanoscale motion of two metal spheres typically 0.1mm and 1.0mm in radii. The smaller sphere is mounted near the end of a single mode fiber optic element that serves both as a cantilever and a monitor of the sphere's motion. A 1.5mW superluminescent diode (SLD) with a center wavelength $\lambda_c$ = 850nm and spectral width $\Delta\lambda$ = 30nm is coupled into the single mode fiber with a numeric aperture N.A. ~ 0.1. The cone of light exiting the end is imaged onto a four-quadrant photodiode that is placed approximately 100μm from the end. The 2-D displacement of the tip in the plane of the photodiode is measured with a sensitivity of $6x10^{-13} m/\sqrt{Hz}$. The details for this method of detection will be discussed in a separate publication [17]. The stiffness of the fiber cantilever is determined by its diameter and by the distance between the metal tip and the point where the fiber is clamped. Experiments between two gold surfaces were performed with two set-ups; in one a 200μm diameter sphere is glued 0.5mm from the free end of a 2.5mm long, 80μm diameter silica fiber; in the other the fiber is125μm in diameter. The respective force constants are $k_f = 80 \ N/m$ and 98N/m for both the transverse and longitudinal displacements. The measurements between gold and platinum used the 98N/m set-up. The larger sphere is brought into contact with the fiber-mounted sphere by means of a 2-D piezo driven stage[18]. Contact is monitored through the mechanical displacement of the fiber along with the change in conductance. A 4-probe conductance measurement is made by biasing the junction with a 0.5VDC potential difference and limiting the current by adding 5kΩ series resistance. The voltage difference across the junction is measured using a specially designed logarithmic amplifier that provides more than four decades of resolution [19].

When contact is made the shift in resonant frequencies of the two fundamental modes yields the longitudinal $k_l$ and transverse $k_t$ stiffness of the junction.

$$(1) \ k_l + k_f = m_{eff}\omega^2$$

Here, $m_{eff} = k_f/\omega_0^2$ is the effective mass of the tip with the attached gold ball and $\omega_0$ is the frequency of the fundamental mode. For the Au-Au and Au-Pt measurements, $\omega_0/2\pi = 4.0\,kHz$. The force constant of the fiber optic cantilever is given by $k_f = 3\pi E_s R^4/4l^3$ [20], where $E_s$ is the Young's modulus of a silica rod, $R$ is the radius of the fiber and $l$ is the distance from the base of the cantilever to the gold ball. In order to generate a response at the resonant frequency a small amount of broad-band noise in the range $3.5\,kHz < f_n < 10\,kHz$ is added to the DC voltage applied to the piezo controlling the tip-sample separation. The RMS response amplitude of the tip at the resonant frequency is 0.4Å. A continuous time record of the longitudinal and transverse displacement is acquired at 20ks/s using a 16 bit digitizer with $1.8 \times 10^{-3}\,\text{Å}/bit$ accuracy. The shift to the fundamental modes of the fiber optic cantilever is measured by breaking the data into 1500 point segments, Fourier transforming each segment and recording the location of the mode as a function of time. Each component of stiffness is calculated from the frequency shift using (1). All measurements were performed after baking the surfaces at 60°C at $10^{-6}$ torr for several days. The use of stacked piezo elements does not permit baking at higher temperatures; therefore, in order to purge water from the system, all surfaces were treated with 1-2 torr of 2,2-dichloropropane vapor introduced into the system during baking [21].

A radius '$r_c$' is assigned to the junction via Wexler's formula [22] which relates the conductance through an aperture to its surface area.

$$(2)\quad G_W = G_S \left(1 + \frac{3\pi}{8}\,\text{K}\,\Gamma(\text{K})\right)^{-1}$$

In the limit of small Knudsen number $\text{K} = r_c/\ell_e$, the expression for the conductance was derived by Sharvin [23].

$$(3)\ G_s = \frac{1}{4} r_c^2\ k_F^2\ G_0$$

Where $k_F$ and $\ell_e$ are the bulk metal Fermi wave number and electron mean free path respectively and $G_0 = 2e^2/h$ is the quantum of conductance[24]. The function $\Gamma(K)$ matches the solutions in the diffusive limit $K >> 1$ with the ballistic limit $K << 1$ and varies slowly with respect to $K$. At the interface of two different metals, $r_c$ is calculated from the measured conductance using

$$(4)\ \frac{1}{G} = \frac{1}{2}\left(\frac{1}{G_W^1} + \frac{1}{G_W^2}\right)$$

where, $G_W^1$ and $G_W^2$ are calculated using (2) for the two metals. If we define the transition between the ohmic and ballistic regions to be where $G_W/G_s = 0.8$, then for Au-Au contacts this occurs at $r_c \approx 10\ nm$ as compared to $r_c \approx 3\ nm$ for Au-Pt contacts.

Figure 3 displays an example of the longitudinal and transverse stiffness that can be resolved with this apparatus. Also shown is the radius of the junction that forms between two gold spheres as a result of forcing them into contact at the rate of 0.8Å/s. The steps in radius are presumably related to jumps in the number of conducting channels [25, 26]. By approximating the junction as a uniform cylinder of length $\ell$ and radius $r_c$, an estimate for the junction stiffness can be obtained from elasticity theory. The longitudinal and transverse stiffness are given by $k_L = \pi E_g r_c/\mu$ and $k_T = k_L/2(1+\nu)$, respectively, where $E_g$ and $\nu$ are the bulk Young's modulus and Poisson's ratio for gold. Values for the aspect ratio $\mu = \ell/r_c$ are suggested to lie in the range 6-20 based on imaging of rod-like gold constrictions using high resolution transmission electron microscopy (HRTEM) [27-31]. The calculated values for these two components of stiffness, taking $\mu = 6.5$ for all values of $r_c$, are displayed in Figure 3 along with the measured data. The overall agreement between measurement and elasticity theory suggests that residual contaminants do not significantly alter the mechanical properties of the junctions.

Measurements of rupture are carried out by first bringing the two spheres into contact then pulling them apart by applying a 9nm/s linear ramp to the z-piezo. The stress plotted in Figure 4 is determined from the conductance and force at the moment of rupture. For the Au-Au measurements, it is observed that the stress that is required to rupture the smallest junctions $r_c \approx 3.2\,nm$ is $\sigma \approx 10\,GPa$, which corresponds to the theoretical limit to rupture a gold lattice[32]. Previous measurements [11, 25, 33] have also reported yield stress in the 3-8 GPa range for gold point contacts with diameter between 1-2 nm. We find that as the radius of the junction grows to 70nm, the stress decreases nearly by a factor of 5 to $\sigma \approx 2\times 10^9\,Pa$. For Au-Pt junctions, the rupture stress drops nearly a factor of 20 as the radius increases from 10nm to 100nm and approaches the tabulated value for macroscopic platinum samples $\sigma = 1.4\times 10^8\,Pa$[34]. The region with a slope of −1.0 can be interpreted as due to failure of a ring of atoms with radius $r_c$. Molecular dynamics simulations have shown that plastic yield for nanometer size junctions involve atoms primarily near the perimeter of the constriction [35], thus the rupture force scales as $r_c$ and the stress scales as $r_c^{-1}$.

Motion of the large sphere transverse to its junction with the small sphere generates a sequence of stick-slip events as shown in Figure 5. The rupture stress and contact radius for these events has also been plotted in Figure 4, where it can be verified that they coincide with the measured forces of fracture. We conclude that for this system stick-slip friction has its origin in the junctions that spontaneously form at the interface of two bodies. The force needed to break the junction is the force required to initiate a slip event. As the time between slips increases, so too does the strength of the junction that forms [Figures 1 and 6]; thus, a greater force is needed to generate the next event. Simulations of friction between Cu surfaces have been made by Sørensen *et al.* [36], who find that neck formation and rupture can give rise to frictional coupling between surfaces. In their simulations, neck formation resulted from simulated annealing. Hence, no comparison can be made with our measured long time scales. The dependence of this timescale on temperature remains to be seen.

Our observations on the tribology of metal surfaces, also applies to incommensurate surfaces. Figure 6 displays the spontaneous growth of a cold-weld between gold and platinum. Here, the ramp applied to the z-piezo is switched off at t = 7 seconds. After this time the surfaces weld at a rate that is more rapid than the Au-Au experiments. Despite the more rapid growth rate observed for these junctions, the rupture stress for a given radius is the same as for the Au-Au junctions.

In conclusion, contact between metallic surfaces leads to a spontaneous plastic deformation that is independent of the effects of the applied load. The deformation is due to bonding which initiates at the angstrom level and proceeds up to the nanometer level on a long time scale. These bonded regions account for the forces of friction. Resolution of the growth, and role, of these junctions in the macroscopic phenomenon of friction was achieved through the use of a newly designed surface probe capable of measuring forces between macroscopic bodies with atomic resolution.

We acknowledge K. Holzcer and J. Gimzewski for valuable discussions regarding the technique and experimental results and H. Lockart for technical assistance in the construction of the device. This work is supported by the U.S. Department of Energy (Division of Engineering Research).

Figure Captions

Figure 1 – Plot of longitudinal displacement and contact radius after two gold surfaces are brought into contact. The z-piezo is displaced until contact is established at t = 0 sec., after which time the piezo voltage is held constant. The radius is calculated from the measured conductance using (2) with $k_F = 1.21 \text{Å}^{-1}$, $\ell_e = 34.2\ nm$ and $K = 0.8$.

Figure 2 – Diagram of experimental setup showing the fiber optic cantilever and detector arrangement. The gold ball attached to the fiber is made by taking a segment of 10μm diameter 99.99% gold wire and bending it in half. The bent section is then melted using a torch and forms a ball approximately 200μm in diameter. The two halves of the gold wire, which are naturally fused to the sphere, are used in the 4-wire conductance measurement. A 250μm gold or platinum wire is heated at one end and forms the larger 2mm diameter ball that is attached to piezo stage. The vector $\vec{g}$ is drawn to show the orientation of the apparatus with respect to gravity.

Figure 3 – Plot of the longitudinal and transverse stiffness of a gold junction that forms during contact. Data is acquired while the surfaces are brought together at 0.8Å/s. The boxes (triangles) are the theoretically determined values of longitudinal (transverse) stiffness.

Figure 4 – Rupture stress vs. contact radius for Au-Au (●) and Au-Pt (□) junctions. The data is taken by bringing the surfaces into contact to form a junction of a given size. The longitudinal displacement and conductance is digitized while a linear ramp applied to the z-piezo pulls the surfaces apart. The stress $\sigma = k_f \Delta z / \pi r_c^2$ is calculated from the maximum longitudinal displacement $\Delta z$ and conductance recorded immediately prior to rupture. The data points represent the average of several hundred such events for each pair of surfaces. The open circles with crosses represent the transverse rupture stress for Au-Au surfaces prior to the slip events shown in Figure 5. Note that the longitudinal and transverse stress at a given contact radius are the same indicating that bond formation is the primary mechanism of frictional coupling for these surfaces.

Figure 5 – Two gold surfaces are brought into contact and a linear ramp is applied to the x-piezo (transverse direction) at 2.7nm/s. The force applied to the junction is calculated by multiplying the displacement by the force constant of the cantilever, which for this measurement is $k_f = 80 \ N/m$. The surfaces remain stuck together until 100s, until the transverse force that builds up in the cantilever is sufficient to rupture the junction. Negative values for the normal force represent compression between the surfaces. After each event, the normal force decreases. As the scan proceeds, the relative transverse displacement between the centers of the spheres increases, thus decreasing the normal displacement. At the end of the scan, the spheres have moved completely out of contact. The points of rupture, indicated by the open circles with crosses, are plotted on the stress curve in Figure 4.

Figure 6 – Time scales for junction growth for Au-Pt junctions. Following contact, the platinum ball is pressed into the gold ball by displacing the cantilever 10nm. The displacement to the z-piezo is then turned off and the growth of the junction radius is measured for the next 120s along with the longitudinal displacement of the tip. As with the data shown in Figure 1, positive displacement values indicate that the centers of the spheres get closer presumably due to one or both of the surfaces flattening as the junction forms. The contact radius is calculated using (3) and (4) with $K = 1.0$. The material constants for platinum are $k_F = 1.18 \text{Å}^{-1}$ and $\ell = 8.7\, nm$. The noise that is visible in both the displacement and radius curves at t = 40 s is due to a short burst of undamped motion. As seen in the inset, this 2.5Hz motion with peak-peak displacement of 15Å (i) causes a corresponding change in the junction radius (ii). Note that for contact radii $r_c \approx 8\text{Å}$, the radius changes in units of $2e^2/h$.

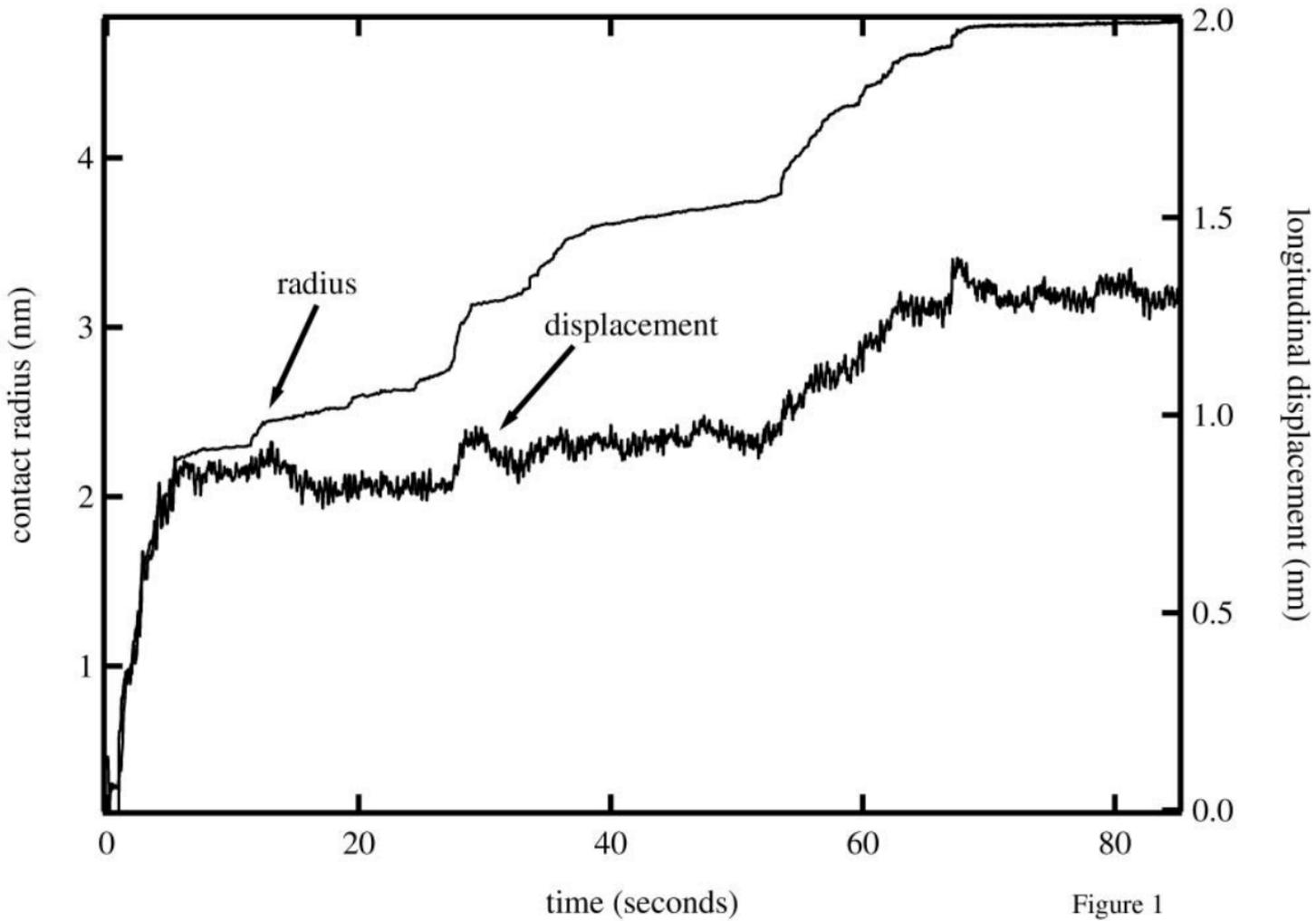

Figure 1

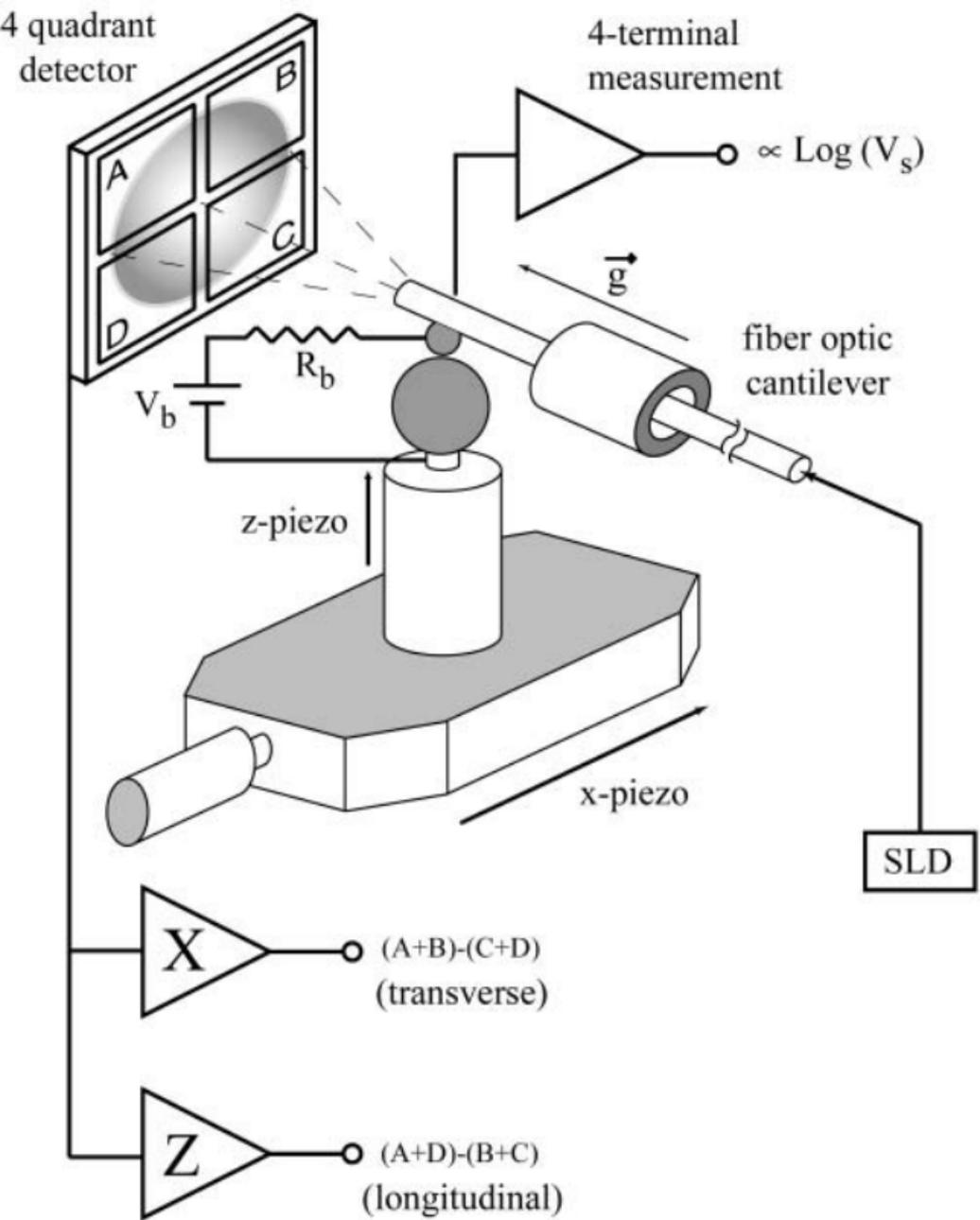

Figure 2

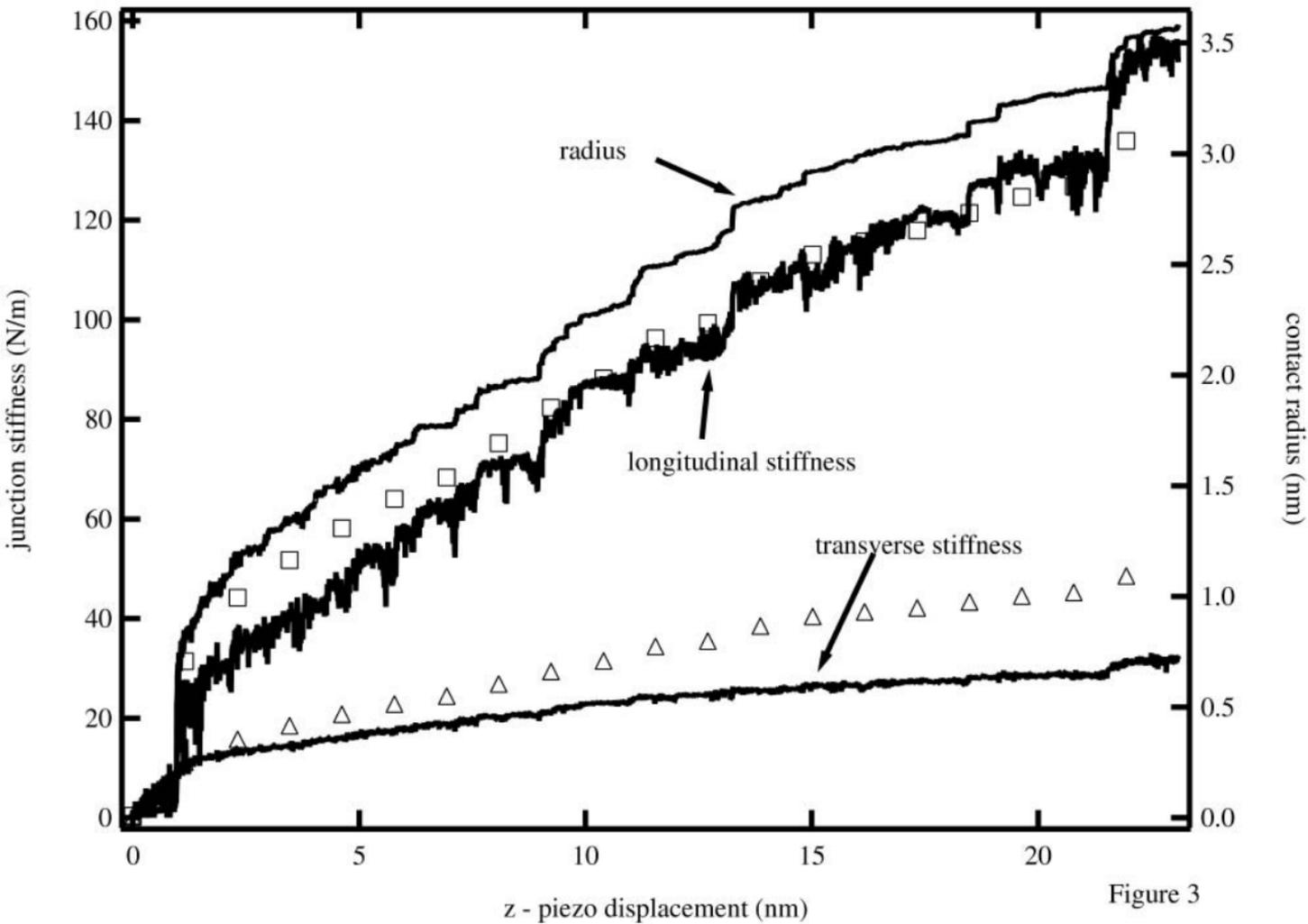

Figure 3

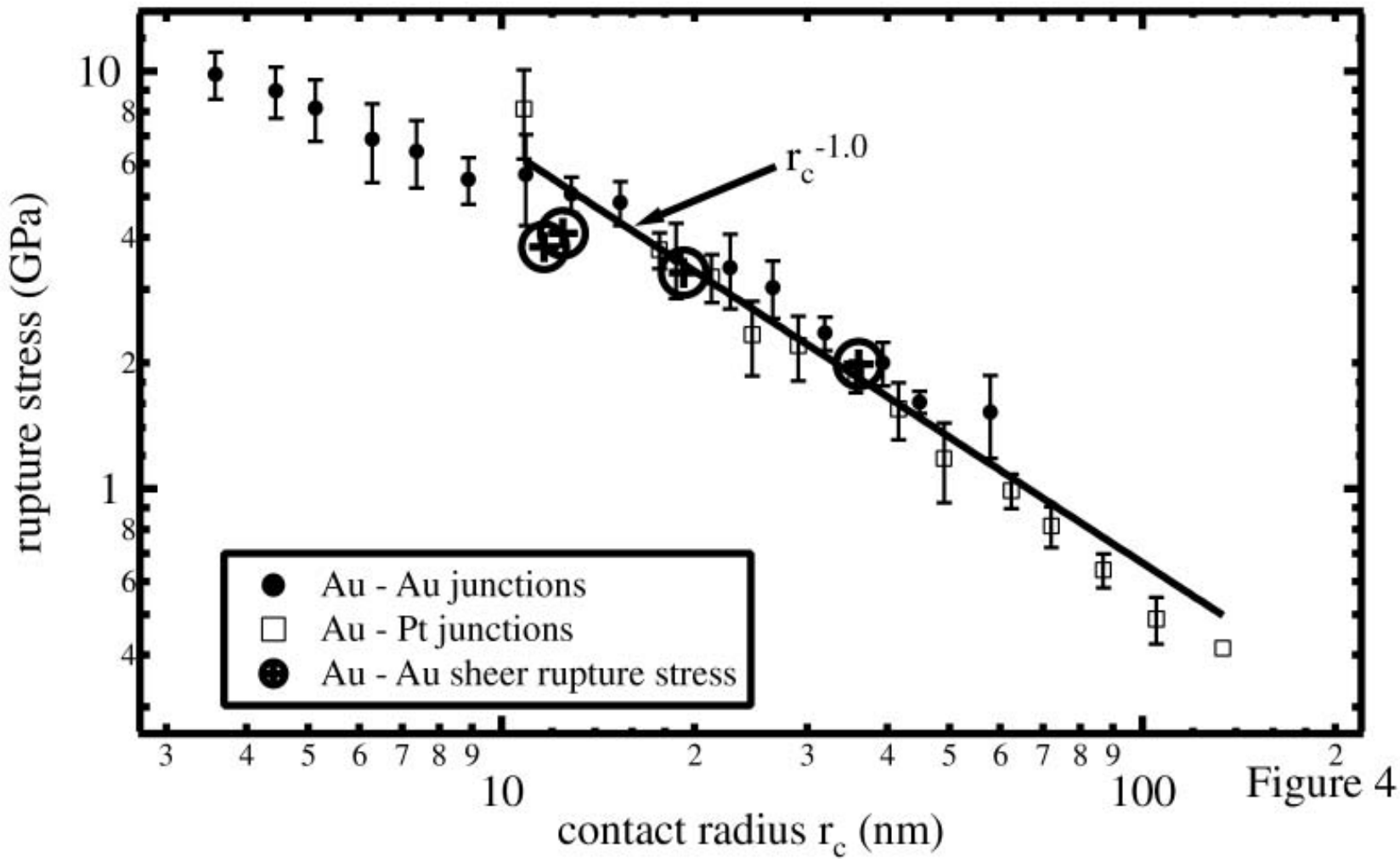

Figure 4

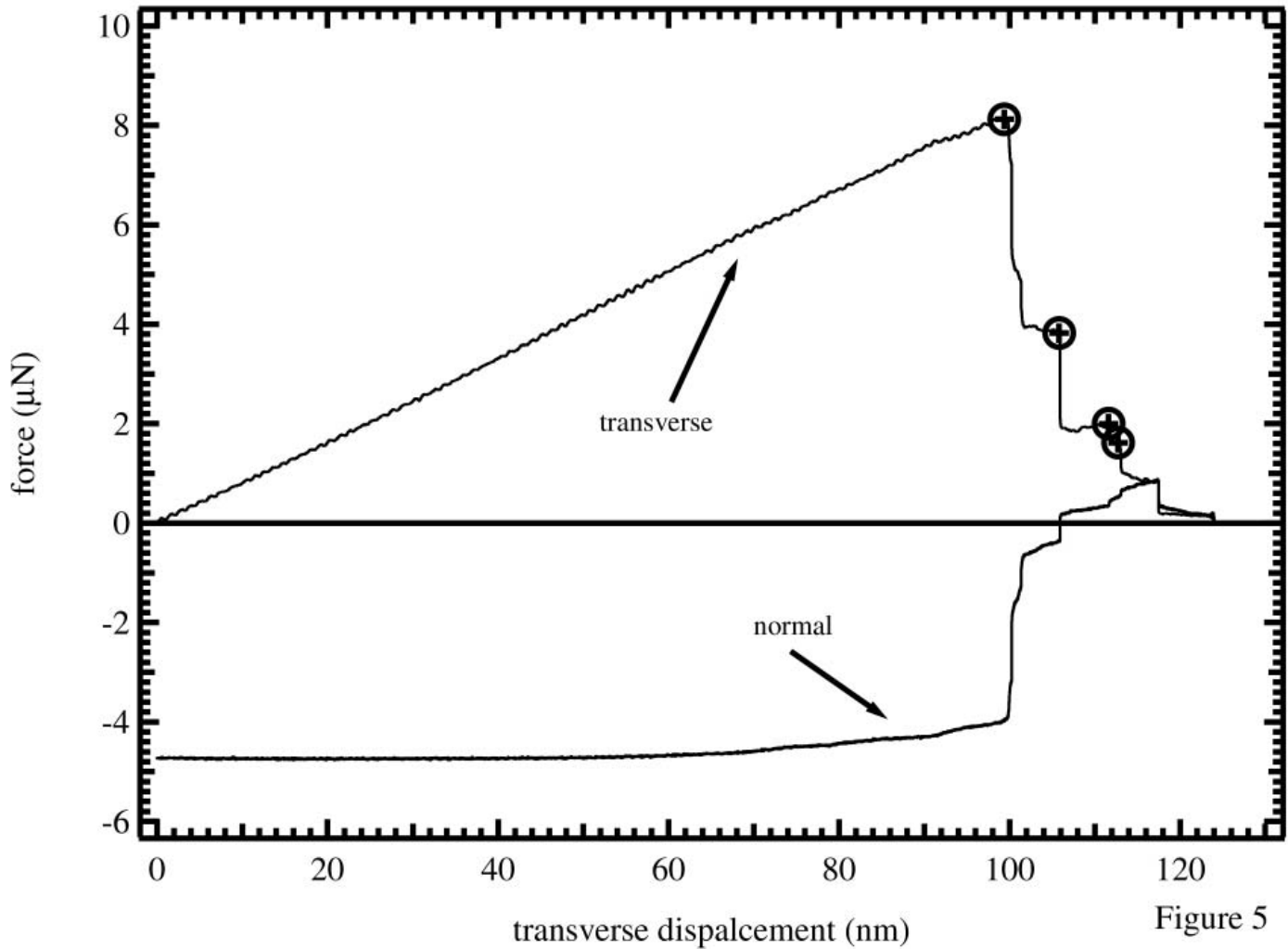

Figure 5

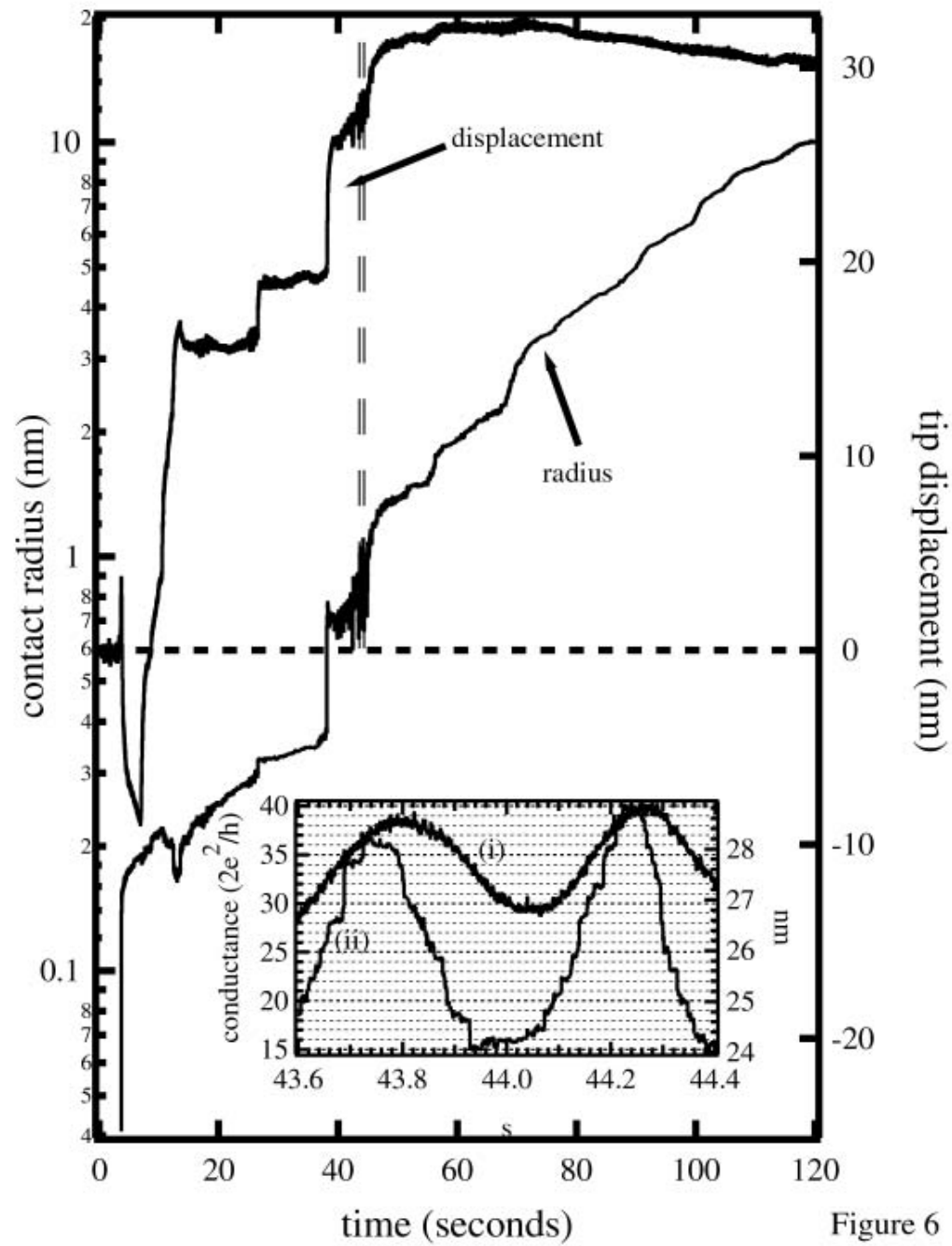

Figure 6